# Switchable plasmonic scattering of nanogaps for linewidth-tunable random lasing

Xiaoyu Shi[1], Zhixing Xu[1], Yaoxing Bian[1], Qing Chang[1], Hongbiao Cui[1], Dahe Liu[1] and Zhaona Wang[1,*]

[1]Applied Optics Beijing Area Major Laboratory, Department of Physics, Beijing Normal University, Beijing, China, 100875

## ABSTRACT

Linewidth-tunable lasers have great application requirements in the fields of high-resolution spectroscopy, optical communications and other industry and scientific research. Here, the switchable plasmonic scattering of the metal particles with plenty of nanogaps is proposed as an effective method to achieve linewidth-tunable random lasers. By using the nonlinear optical effect of the environment medium, the metal particles demonstrate the transition from local scattering of nanogaps with high spatial frequency to traditional Mie scattering free from detail information with increasing the pump power density. Based on these two scattering processes, random lasers can be continuously driven from a narrow-linewidth configuration exhibiting nanogap effect dominated resonances to a broad-linewidth regime of collectively coupling oscillating among nanowires (or nanoflowers), demonstrating the dynamic range of linewidth exceeds two orders of magnitude. This phenomenon may provide a platform for further studying of the conclusive mechanism of random lasing and supply a new approach to tune the linewidth of random lasers for further applications in high-illumination imaging and biology detection.

**Keywords**：Random lasers; Linewidth-tunable; nanogap effect; Au-Ag nanowires

# Introduction

Linewidth is a pivotal parameter for lasing that characterizes the temporal coherence of light sources, which is generally related to the coherent time $t_c$ with the relation of $\Delta \upsilon = 1/t_c$[1-3]. The narrower linewidth $\Delta\lambda$ or $\Delta\nu$ of light source, the better temporal coherence is. In the conventional cavity lasers, the linewidth has been optimized and regulated for their huge and extensive applications in several fields of high-resolution spectroscopy[4,5], optical communications[6,7] and other industry and scientific research. As a new type of cavity-free lasers, random lasers (RL) are built by multi-scattering of light and have demonstrated several potential applications in the fields of medical diagnosis and sensing[8,9], stretchable optoelectronic device[10,11] and speckle-free bioimaging[12] according to its characteristics of easy-fabrication, flexiblity and low-spatial coherence. Tuning linewidth of random lasers is a key role to promote their practical applications in high-illumination imaging and biology detection. However, the cavity-free disorder structures set a hurdle for tuning linewidth for the absence of a stationary cavity. How to realizing linewidth-tunable random lasers is a desirable but challenging issue.

Several attempts have been proposed to changing the linewidth of random lasing. Historically, random lasing with an emission spectral width of few nanometers is attributed to the incoherent feedback[13-17]. And the narrower ones with sub-nanometer bandwidth are looked as the result of coherent feedback[8,18-22]. By using these two types of feedback mechanism, the linewidth of random lasing was poorly tuned by simply modifying the composition information of random lasers as the concentrations of scatterers[21] and gain[23]. Furthermore, a relative flexible approach has been proposed by changing the pumping conditions[24,25]. The linewidth of random laser is thus tuned by utilizing different pumping source with different pulse duration or the special configuration of pumping light. Additionally, nonlinear optical interaction in random medium has been utilized to vary and/or tune the temporal profile and spectral characteristic of random lasing. Professor R. Vallée has demonstrated the competition relations between two coexistence nonlinear effects of random lasing and stimulated

Raman scattering. These nonlinear effects have introduced different spectral emissions with a bandwidth of sub-nanometer or several nanometers[26, 27]. Prof. Cao has reported that the third-order nonlinearity can modify the laser emission intensity and laser pulse width[28]. These linewidth-tunable approaches have demonstrated some potential value in linewidth-tunable random lasers. However, relative small tuning range of linewidth and high requirements in advanced devices hinder their further developing and practical applications. Therefore, it is necessary to find an easy-to-implement linewidth-tunable method for random lasers with large tuning range.

In this work, a linewidth-tunable random laser is obtained based on the switchable plasmonic scattering of nanogaps by utilizing the nonlinear optical effect of the environment medium within and near the gold-silver bimetallic porous nanowires (Au-Ag NWs). With increasing the pump power density, the scattering of the metal particles with plenty of nanogaps switches from local scattering of nanogaps to the contour profile scattering without detail information. And the random lasers are naturally driven from a narrow-linewidth configuration induced by nanogap effect to a large-linewidth regime when turning off nanogap effect. This phenomenon is observed in different random systems with different gain materials and/or different nanogap-based scatterers. The results may supply a new approach to achieve the linewidth-tunable random lasers for their further applications in high-illumination imaging and biology detection.

## Results and discussions

Different from the cavity lasers, random lasers are highly dependent on the multi-scattering processes of the photon in the random medium. And the scatterer is an important part of random lasers. To improve the performance of random lasers, varies of metal nanoparticles have been introduced in the random systems replacing dielectric nanoparticles. Generally, there are two kinds of scatterers. One is the metal nanoparticle with smooth surface that supplies enough gain by coupling the surface plasmonic resonance of single particle and multi-scattering (Figure 1a) [29, 30]. The other is the metal particle with plenty of nanogaps and/or nanotips of concave-convex structure[16, 31]. These nano-singularities could generate hotspot effect that greatly enhances the local electromagnetic field. By combining the hotspot (nanogap) effect within the particles and the multi-scattering between different scatterers, high performance random lasing is obtained when the total gain is larger than the loss in a disorder system (Figure 1b) [32, 33]. To better utilize the hotspot effect, the second type of random systems in Figure 1b are fabricated by dispersing Au-Ag NWs with lots of nanogaps (Supplementary Figure S1a) in a Rhodamine 6G (R6G) solution (Supplementary Figure S1b). The concentration of Au-Ag NWs is $1.09 \times 10^8$ ml$^{-1}$ with a mean free path of the system estimated to be $l^* > 2$ cm[18,19], indicating that the system is a weakly scattering system as a whole.

Figure 2a presents the evolution of the emission spectra from random lasers with a R6G concentration of 1.67 mM under different pump power densities. At a lower pump power density of 0.16 MW cm$^{-2}$, the spectrum exhibits a broad spontaneous emission with a maximum at 582 nm and a full width at half maximum (FWHM) of 60 nm. When the pump power density exceeds the threshold of 0.33 MW cm$^{-2}$ (Figure 2e and 2g), the emission spectra displays several peaks with sub-nanometer bandwidth. By using a high-resolution spectrometer, number of sharp spikes with the FWHM of around 0.03 nm are observed in Figure 2d when random laser is pumped at a power density of 0.78 MW cm$^{-2}$, demonstrating an extremely narrow-linewidth state of RL-I. In this state, the corresponding Q factor of these random lasing modes is about 19460 which is comparable to traditional lasing systems[34, 35]. As the pump power density further increases, the spikes disappear gradually with the spectral peak blue-shifted as shown in Figure 2a and the high-resolution spectra in Figure 2c. There is only one smooth peak with FWHM of several nanometers marked as RL-II, corresponding to a low Q factor of ~ 90. However, while the pump power density continues to increase above 7.24 MW cm$^{-2}$, emission lasing demonstrates few peaks with bandwidths of about 0.8 nm in low-resolution spectra in Figure 2a, as well as in high-resolution spectra shown in Figure 2b. And the corresponding Q factor is about 720. This emission is different from the one in Figure 2c and is marked as RL-III. The evolution of intensity and linewidth for the random lasing mode at 578.49 nm are shown in Figure 2e, showing three inflection points and corresponding three different types of resonance regions. Beyond the threshold of 0.33 MW cm$^{-2}$ (Figure 2e and 2f), the intensity of random lasing at 578.49 nm increases rapidly with the linewidth decreasing to sub-nanometer (RL-I in Figure 2e). When the pump power density surpasses 2.29 MW cm$^{-2}$ (RL-II in Figure 2e), the intensity begins to decrease and the linewidth increases to several nanometers. And when the power density is larger than 22.18 MW cm$^{-2}$ (Figure 2e and 2g), the intensity increases slightly again with the linewidth decreasing to a sub-nanometer (RL-III). Furthermore, these three types of random lasing show good stability (Supplementary Figure S2).

To further reveal the dynamic performance of these random lasing modes in different regions, their temporal properties are statistically studied and demonstrated in Figure 3 and S3. Notably, the random lasing modes in the region of RL-I are changeable (Supplementary S3a), while the modes for RL-II and RL-III are relative repeatable from shot to shot (Supplementary Figure S3b and S3c). The intensity values of the mode at 582.12 nm show an intense fluctuation of $I_{max}/I_{min}=15$ under a pump power density of 0.78 MW cm$^{-2}$ in the narrow-linewidth state of RL-I (Figure 3a). However, the intensity fluctuation ratio of the random lasing modes is $I_{max}/I_{min}=1.4$ under 3.54 MW cm$^{-2}$ in the broad-linewidth state of RL-II (Figure 3b) and $I_{max}/I_{min}=1.7$ under 22.1 MW cm$^{-2}$ in the broad-linewidth state of RL-III (Figure 3c), respectively. The inter-mode spectral correlations of the nanogap-based RL are illustrated in Figure 3d-f through recording the intensity of the two random lasing modes at different time and under some pump power densities. The random distributed points demonstrate that the random lasing modes at 582.12 nm and 579.98 nm are uncorrelated in the extremely narrow-linewidth state of RL-I. While the random lasing modes are strongly correlated in the broad-linewidth states of RL-II and RL-III.

The linewidth-tuning mechanism of the nanogap-based random lasers is shown in Figure 4a and 4b attributing to the switchable plasmonic scattering of nanogaps modulated by nonlinear effect of the environment. Under a lower pump power density, nanogaps with different separation spacing greatly enhance the local electromagnetic field (Figure 4a). And the nanogaps with smaller spacing support the larger field enhancement factor and stronger local electrical field, meaning larger gain in the random system. When the gain is larger than the loss, the random lasing with narrow-linewidth is observed by turning on the nanogap effect whose local scattering characteristics lead to the uncorrelated properties of random lasing modes characterized as RL-I. Also, enough strong local electromagnetic field may introduce the nonlinear effect of the surrounding medium, leading to varying the local refraction index. As the pump power density increases, more nanogaps can supply enough strong local electromagnetic field to induce a refraction index gradient domain near the nanogaps (Figure 4b). Different domains are generally separated as shown in Figure 4d and supplementary Figure S4. And the scattering coefficient of Au-Ag NWs in ethanol solution (Figure 4c) illustrates an increasement with the pump power density. This scattering mode is called nanogap-based scattering. When the pump power is stronger, each refraction index gradient domain becomes larger and different domains are coupled each other to form a new scattering surface simulated as the supplementary Figure S4. This new surface is the tangential surface to all of these secondary scattering of different domains, just as Huygens' principle giving the wave-front.[36] And the Au-Ag NW is then packed in a capsule of refraction index gradient shell as shown in Figure 4b. The shell can turn off the nanogap scattering and the nanowires can scatter the light wave as a traditional nanowire with the smooth surface. This state of turning off the nanogap scattering introduces a relative large loss factor for the random system. The broad-linewidth random lasing with a strong correlation is thus induced as demonstrated in the regions of RL-II and RL-III. Due to the shielding effect of the shell, the scattering coefficient maintains a relative stable value as the pump power densities increasing. When the pump power density is large enough, Raman signal of R6G can work as a seed and obtain enough gain through multiple scattering process to form a stimulated emission in the RL-III regime.[37-40] The FWHM of random lasing mode decreases again. This is the fundamental of tuning the linewidth of random lasing. To further explore nonlinear plasmonic scattering property, the spatial intensity distribution near an Au-Ag NW in dye liquid is shown in Figure 4d by using a home-made dark-field microscope (Supplementary Figure S5). The bright spots around the nanogaps within Au-Ag NWs increase as the pump power increasing for the nanogap effect. Until the pump power is intense enough, a bright layer is formed surrounding the nanowire, shielding the nanogap effect.

The robustness of the linewidth-tunable approach is systematically investigated in Figure 5 by varying the dye concentration of $C_{R6G}$ and replacing R6G. For the nanogap-based random lasers, transition from narrow-linewidth state of RL-I to broad-linewidth state of RL-II operates well in a wide range of R6G concentration between 0.067 mM and 2 mM. When $C_{R6G}$ is decreased to 0.313 mM, there is no broad-linewidth state of RL-III emerged even when the pump power density is large enough (Figure 5b and Supplementary Figure S6). This is because $C_{R6G}$ is too small to provide enough gain to form the Raman based random lasing. On the contrary, when $C_{R6G}$ is increased to 3.13 mM, there is no narrow-linewidth state of RL-I due to a smaller threshold of Raman based random lasing relative to that of random lasing (Supplementary Figure S7). More importantly, the similar linewidth-tunable behavior from narrow-linewidth state to broad-linewidth state has been obtained in nanogap-based random systems mixed with Coumarin 440 (C440) (Figure 5c) or 4 - (dicyanomethylene) – 2 – tert – buty l – 6 - (1,1,7,7 – tetramethyljulolidin – 4 – yl - vinyl) - 4h - pyran (DCJTB) (Figure 5d). The results verify that the proposed linewidth-tunable approach works well for the random lasers with different gain concentrations and even different gain materials.

Furthely, the extendibility of the suggested approach is carefully demonstrated in Figure 6 by varying the scatterer concentration and/even replacing Au-Ag NWs by other nanogap-based particles. Three regional transformation of RL-I, RL-II and RL-III is clearly observed in the random systems in a very wide concentration range of Au-Ag NWs from $1.21 \times 10^7$ ml$^{-1}$ to $3.87 \times 10^8$ ml$^{-1}$ (Supplementary Figure S8a), meaning linewidth easily tuned. Besides, the phenomenon of tuning from narrow-linewidth state to broad-linewidth state by changing the pump power density is also observed in the nanogap-based random systems with silver nanoflowers (Ag NFs) (Figure 6 and Supplementary Figure S9), indicating the ubiquity of this phenomenon. Similarly, the emission spectrum exhibits a narrow-linewidth state with a FWHM of 0.03 nm at the lower pump power densities. As the pump power density increases, the spikes disappear gradually and the random laser is in a broad-linewidth state of RL-II. When the pump power density continues to increase above 11.4 MW cm$^{-2}$, FWHM of random lasing down to about 0.8 nm, meaning a broad-linewidth state of RL-III.

## Conclusions

These linewidth-tunable nanogap-based random lasers possess novel, unique and complement features in comparison with conventional linewidth-tunable approaches based on varying the composition information of random lasers and the pumping conditions. First, the overlooked nonlinear effect of surrounding medium induced by strong pulse is naturally employed to tune the linewidth of random lasing. This is a fundamentally new mechanism which relies on the naturally-existing phenomenon of light induced refractive index variation. Compared with traditional methods to tune the linewidth of random lasing, the proposed approach demonstrated in this work is easier and lower-cost. Moreover, this approach can be extended to other gain materials, including visible and/or infrared luminous materials, and thus broaden the spectra ranges of random lasing for more practical applications. Second, the scatterers with plenty of nanogaps are easily fabricated by the one-step liquid phase synthesis method with an extremely low cost. Easy-fabrication of nanogap materials guarantee the suggested linewidth-tunable approach wide applications. Third, the random laser operates with a relative low threshold of 0.33 MW cm$^{-2}$. In addition, the linewidth can be driven from a small value of 0.03 nm to a large value of 6.5 nm, demonstrating the linewidth tuning range exceeding two orders of magnitude. These unique features of the nanogap-based random laser open the door for revealing the conclusive mechanism of random lasing and supply a new approach to tune the linewidth of random lasers for further applications in high-illumination imaging and biology detection.

## Methods:

### *Preparation of nanogap-based random systems*

The Ag nanowires are synthesized via a polyvinyl pyrrolidone (PVP) assisted reaction in ethylene glycol[41, 42]. The Au-Ag NWs are synthesized through etching the surface of Ag nanowires (Ag NWs) via a galvanic replacement reaction between Ag and $HAuCl_4$ with molar ratio of $\gamma_M = M_{Au}: M_{Ag} = 0.072$ as described in our previous work[16, 43, 44]. The Ag NFs are fabricated by a rapid one-step solution-phase synthesis method[45]. The random systems are built up by mixing the Au-Ag NWs (or Ag NFs) suspension and gain medium of R6G (C440, DCJTB) with different ratios to form different nanogap-based random lasers.

### *Materials Characterization*

The microscopic structures of the Au-Ag NWs and Ag NFs are characterized by SEM (Hitachi S4800 microscope). A dark-field microscope system is used to detect the intensity distribution of Au-Ag NWs in R6G solution.

### *Optical measurements*

Mixture of R6G and Au-Ag NWs suspension is dropped into a cuvette with a length of 20 mm, width of 10 mm, and a height of 45 mm. For the photoluminescence spectral measurements, the 532 nm pulse with a pulse duration of 8 ns and a repetition rate of 10 Hz (Continuum model Powerlite Precision 8000) is used as a pumping source for different nanogap-based random lasers with R6G (or C440). The third harmonic generation of the Q-switched Nd:YAG pulsed laser at 355 nm is used to pump nanogap-based random lasers with C440. Emission light is collected by using an optical fiber spectrometer (Ocean Optics model Maya Pro 2000) with a spectral resolution of 0.4 nm or a high-resolution spectrograph with an intensified charge-coupled device (ICCD, Princeton Instruments Acton SP2750).

ASSOCIATED CONTENT

**Supporting Information**

A: The set-up of random lasing systems

B: Stability of nanogap-based random lasers

C: Dynamic properties of the random lasing modes in different states

D: Set-up of dark-field microscope

E: Electric-field distribution near the nanogaps under different pump energy

F: Linewidth-tunable random lasers with different dye concentrations

G: Linewidth-tunable random lasers with different Au-Ag NWs concentrations

H: Linewidth-tunable regime in a random system based on Ag nanoflowers

AUTHOR INFORMATION

**Corresponding Author**

*E-mail: zhnwang@bnu.edu.cn

**Present Addresses**

[1] Applied Optics Beijing Area Major Laboratory, Department of Physics, Beijing Normal University, Beijing 100875, China.

**Author Contributions**

Z. N. W. and X. Y. S. conceived the idea and designed the experiments. X. Y. S. and H. B. C. performed the experiments. Z. N. W. and X. Y. S. analyzed the data. Z. X. X., X. Y. S. and Y. X. B. characterized the materials. Z. N. W. and X. Y. S. wrote the paper. All authors discussed the results and commented on the manuscript.

**Notes**

The authors declare no competing financial interest.

ACKNOWLEDGMENT

The authors thank the National Natural Science Foundation of China (grant Nos. 11574033, 61275130 and 11074024), Beijing Higher Education Young Elite Teacher Project and the Fundamental Research Funds for the Central Universities for financial support.

**Figure Captions:**

**Figure 1**

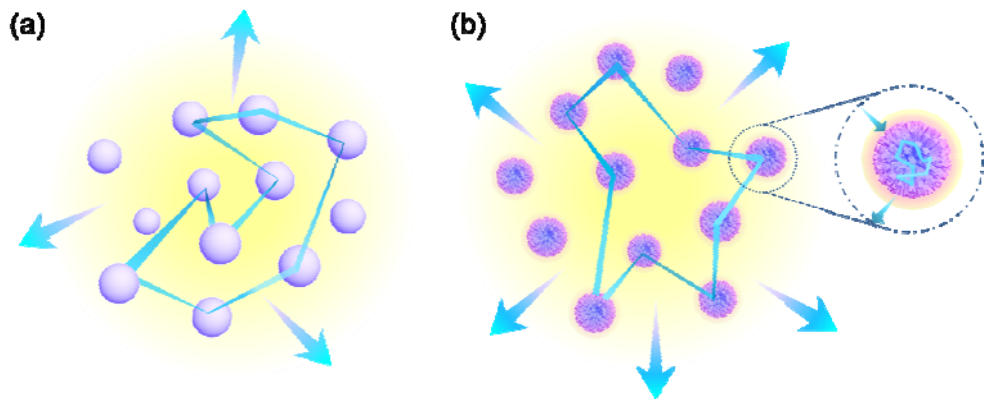

**Figure 1 The schematic diagram of the scattering model for random lasers.** (a)Traditional nanoparticles with smooth surface, and (b)nanoparticles with plenty of nanogaps as the scatterers.

**Figure 2**

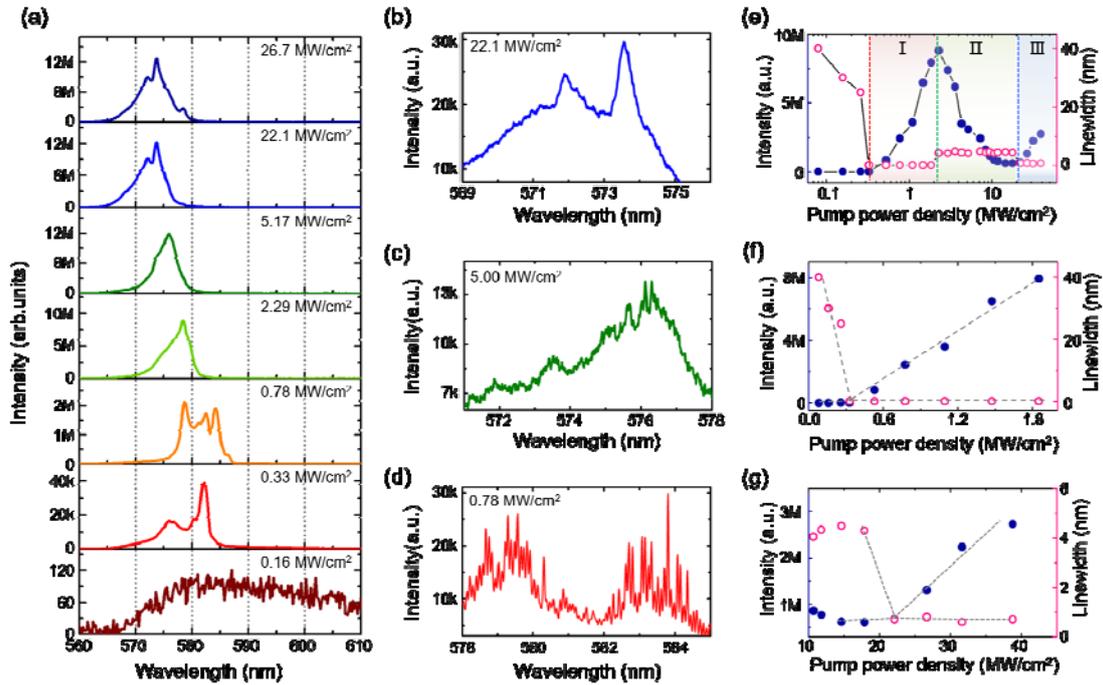

**Figure 2 Spectral characteristics of nanogap-based random systems.** (a) Emission spectra of the random system ($C_{R6G}$ = 1.67 mM) obtained at different pump power densities. (b-d) Specifics of emission spectra obtained at the pump power densities of 0.78 MW cm$^{-2}$ (b), 5.00 MW cm$^{-2}$ (c) and 22.1 MW cm$^{-2}$ (d), respectively. (e) The threshold behavior of random lasing at 578.49 nm in the nanogap-based random system. (f) Magnification of left segment of **e**. (g) Magnification of right segment of **e**.

**Figure 3**

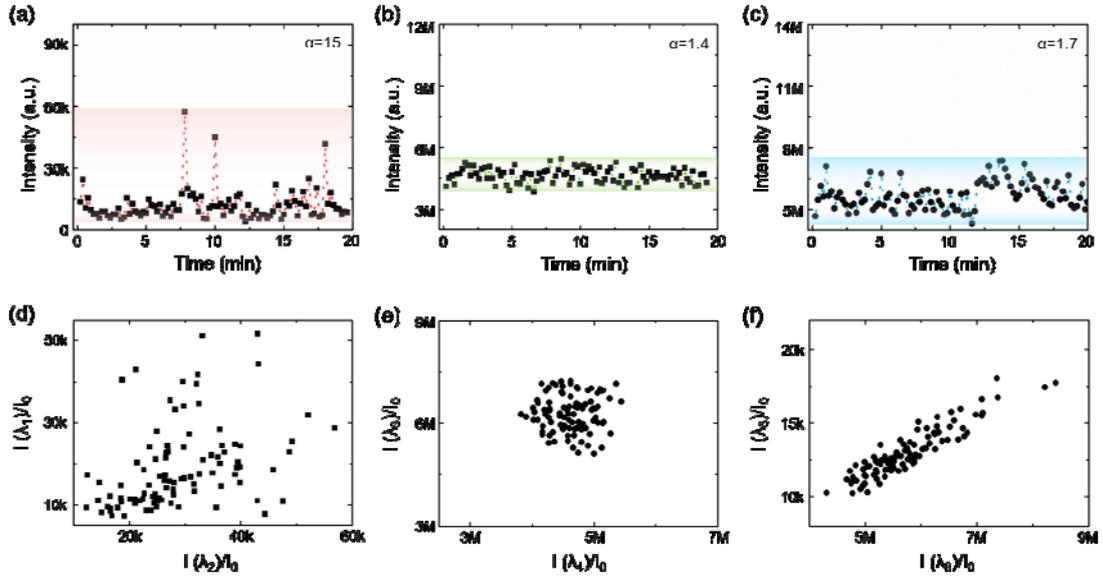

**Figure 3 Dynamic performance of the random lasing modes in different regions.** (a-c) Phase diagrams of maximum intensity values of the multiple modes at the pump power densities of (a) 0.78 MW cm$^{-2}$ ($\lambda$ = 582.12 nm), (b) 3.54 MW cm$^{-2}$ ($\lambda$ = 572.06 nm), and (c) 22.1 MW cm$^{-2}$ ($\lambda$ = 572.06 nm), respectively. (d-f) Phase diagrams of maximum relative intensity values of the random lasing modes at the pump power densities of (d) 0.78 MW cm$^{-2}$ ($\lambda_1$ = 582.12 nm, $\lambda_2$ = 579.98), (e) 3.54 MW cm$^{-2}$ ($\lambda_3$ = 572.06 nm, $\lambda_4$ = 573.77 nm), and (f) 22.1 MW cm$^{-2}$ ($\lambda_5$ = 572.06 nm, $\lambda_6$ = 573.77 nm), respectively.

Figure 4

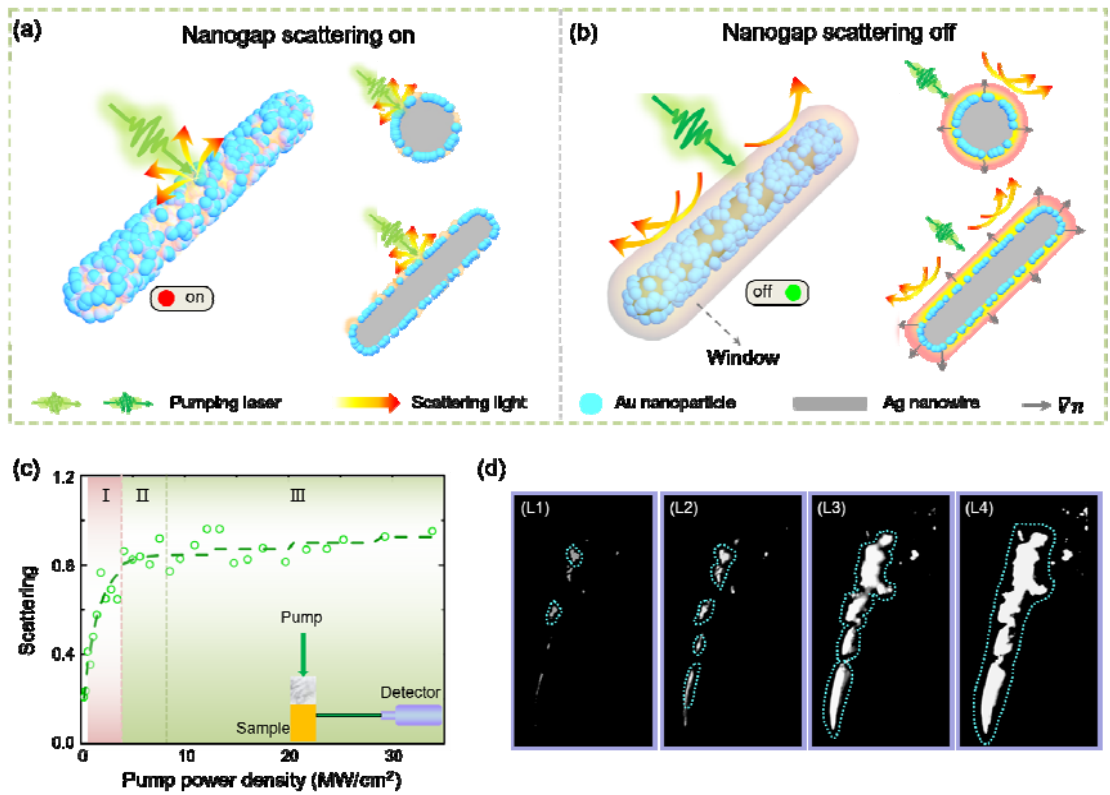

**Figure 4 Linewidth-tuning mechanism of the nanogap-based random lasers**. (a,b) The schematic diagram of the switchable plasmonic scattering induced by the nonlinear optical effect of the medium under a low pump power density (a) and a higher pump power density (b). (c) The nonlinear scattering of Au-Ag NWs under the excitation of 532 nm pulse laser. (d) Spatial distribution of scattering intensity for Au-Ag NWs as the intensity of incident light increases from L1 (3 mW) to L4 (614 mW).

**Figure 5**

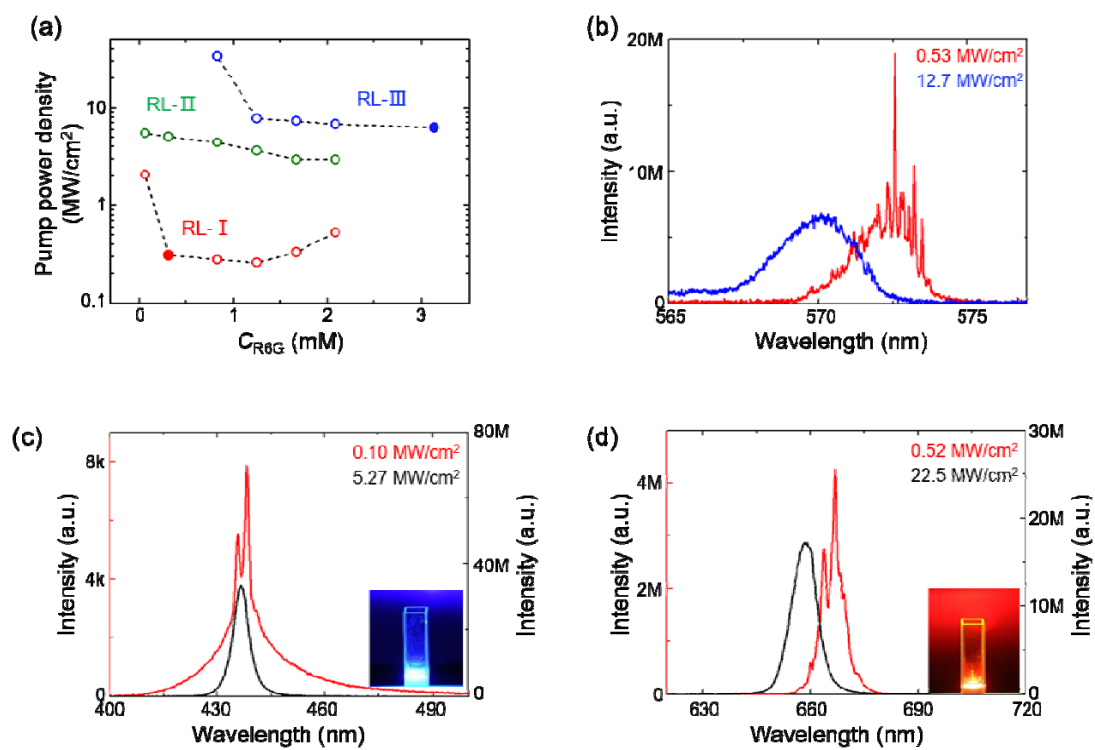

**Figure 5 Extendability of linewidth-tuning approach.** (a) Dependence of $C_{R6G}$ on the threshold and transition points keeping the concentrations of Au-Ag NWs unchanged. (b) The specifics of emission spectra for the sample when $C_{R6G}$ is 0.313 mM. (c) The emission spectra for the sample mixed with C440 and Au-Ag NWs pumped by 355 nm laser. (d) The emission spectra for the sample mixed with DCJTB and Au-Ag NWs pumped by 532 nm laser.

**Figure 6**

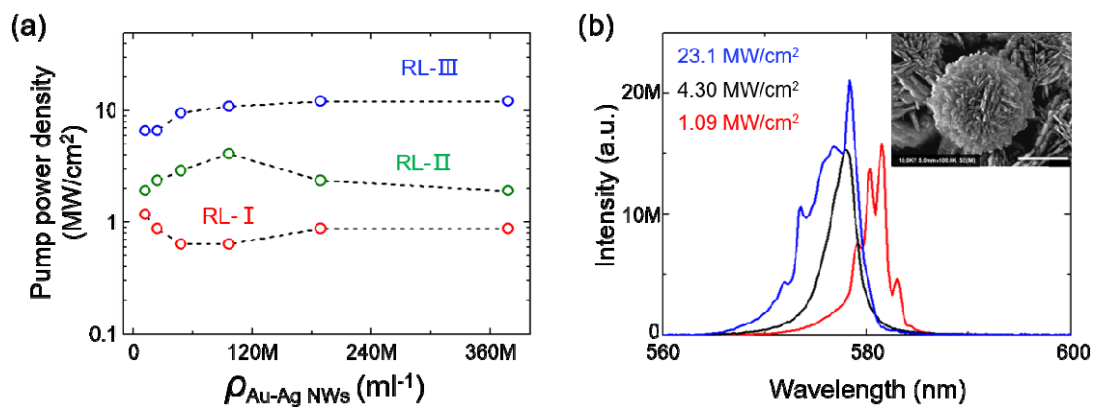

**Figure 6 Linewidth-tuning random lasing based on different scatters**. (a) Dependence of the threshold and transition points on the concentration of Au-Ag NWs. (b) The emission spectra of the Ag nanoflowers based random laser mixed with R6G ($C_{R6G}$ = 0.313 mM). Inset: SEM picture of the Ag nanoflowers.